\documentclass[doublecol]{epl2} 
\usepackage{graphicx,epsfig,epsf}% Include figure files
\usepackage{dcolumn}% Align table columns on decimal point
\usepackage{bm,psfrag,color,amsmath,amssymb}% bold math
\usepackage[matrix,frame,arrow]{xypic}
\title{Ultimate quantum bounds on mass measurements with a
  nano-mechanical resonator}
\author{Daniel Braun\inst{1,2}}
\institute{                    
  \inst{1} Universit\'e de Toulouse, UPS, Laboratoire
de Physique Th\'eorique (IRSAMC), F-31062 Toulouse, France\\
  \inst{2} CNRS, LPT (IRSAMC), F-31062 Toulouse, France
}
\pacs{85.85.+j}{Nanoelectromechanical systems}
\pacs{07.75.+h}{Mass spectrometers}
\pacs{03.67.-a}{Quantum information}

\abstract{I
  establish the fundamental lower bound on the mass that can 
  be measured with a nano-mechanical resonator in a given 
  quantum state based on the fundamental quantum Cram\'er--Rao bound, and
  identify the quantum states which will allow the largest sensitivity for
  a given maximum energy.  I show that with existing carbon nanotube
  resonators it should be possible 
  in principle to measure a thousandth of the mass of an electron,
  and future improvements might allow to reach a regime where one can
  measure the relativistic change of mass due to absorption of a single
  photon, or the creation of a chemical bond.}

\begin{document}
\def\vec#1{\mathbf{#1}}
\def\ket#1{|#1\rangle}
\def\bra#1{\langle#1|}
\def\ketbra#1{|#1\rangle\langle#1|}
\def\braket#1{\langle#1|#1\rangle}
\def\idmat{\mathbf{1}}
\def\caln{\mathcal{N}}
\def\calc{\mathcal{C}}
\def\rhon{\rho_{\mathcal{N}}}
\def\rhoc{\rho_{\mathcal{C}}}
\def\tr{\mathrm{tr}}
\def\bfu{\mathbf{u}}
\def\bfmu{\mbox{\boldmath$\mu$}}

\newcommand{\ri}{{\rm i}}
\newcommand{\re}{{\rm e}}
\newcommand{\bb}{{\bf b}}
\newcommand{\bc}{{\bf c}}
\newcommand{\bx}{{\bf x}}
\newcommand{\bz}{{\bf z}}
\newcommand{\by}{{\bf y}}
\newcommand{\bv}{{\bf v}}
\newcommand{\bd}{{\bf d}}
\newcommand{\br}{{\bf r}}
\newcommand{\bk}{{\bf k}}
\newcommand{\bA}{{\bf A}}
\newcommand{\bE}{{\bf E}}
\newcommand{\bF}{{\bf F}}
\newcommand{\bR}{{\bf R}}
\newcommand{\bM}{{\bf M}}
\newcommand{\bn}{{\bf n}}
\newcommand{\bs}{{\bf s}}
\newcommand{\tbs}{\tilde{\bf s}}
\newcommand{\rSi}{{\rm Si}}
\newcommand{\dB}{d_{\rm Bures}}
\newcommand{\beps}{\mbox{\boldmath{$\epsilon$}}}
\newcommand{\bthe}{\mbox{\boldmath{$\theta$}}}
\newcommand{\blam}{\mbox{\boldmath{$\lambda$}}}
\newcommand{\rg}{{\rm g}}
\newcommand{\xmax}{x_{\rm max}}
\newcommand{\ra}{{\rm a}}
\newcommand{\rx}{{\rm x}}
\newcommand{\rs}{{\rm s}}
\newcommand{\rP}{{\rm P}}
\newcommand{\up}{\uparrow}
\newcommand{\down}{\downarrow}
\newcommand{\hc}{H_{\rm cond}}
\newcommand{\kb}{k_{\rm B}}
\newcommand{\cI}{{\cal I}}
\newcommand{\tit}{\tilde{t}}
\newcommand{\cE}{{\cal E}}
\newcommand{\cC}{{\cal C}}
\newcommand{\Ubs}{U_{\rm BS}}
\newcommand{\qq}{{\bf ???}}

\maketitle

\section{Introduction}
High-quality nano-mechanical resonators can act as extremely sensitive
sensors of adsorbed material.
Impressive progress has been made in this direction over the last few
years: In 2004, experiments reached a level of sensitivity of
femto-grams \cite{lavrik_cantilever_2004}, atto-grams 
 \cite{ilic_attogram_2004}, and two years later already zepto-grams
 \cite{yang_zeptogram-scale_2006}. Gas 
 chromatography at the single molecular  
level was achieved a year ago
\cite{naika_k_towards_2009}, and
brought a vast range of chemical and biological applications in
reach. A mass sensitivity as small as half a gold atom has
been demonstrated
using a nano-mechanical resonator based on a carbon nano-tube
\cite{jensen_atomic-resolution_2008}. At the same time, large efforts
have been spent to cool down 
a nano-mechanical resonator to its ground state, with the ultimate goal 
of engineering arbitrary quantum states (see 
e.g.~\cite{martin_ground-state_2004,rocheleau_preparation_2010,buks_quantum_2008,Physics.2.40,Kippenberg08}). 
The ground state  
was reached very recently for a 
piezo-electrical device \cite{oconnell_quantum_2010}.  It is 
therefore natural to ask whether the 
sensitivity of mass measurements could be increased further by engineering
the quantum state of a nano-mechanical resonator, and what would
be the truly fundamental lower bound on the mass that can be measured based
only on the laws of quantum mechanics. Early 
on, 
theoretical investigations tried to find the limitations of mass
measurements with a nano-mechanical resonator
\cite{cleland_noise_2002,clerk_quantum-limited_2004,giscard_quantum_2009}.
But the bounds which were derived so far
assume that one measures the linear response of the resonator
driven at its resonance frequency
\cite{cleland_noise_2002,clerk_quantum-limited_2004,ekinci_ultimate_2004,giscard_quantum_2009}.  
In the experiments, a variety of different
read-out and/or cooling techniques (e.g.~optical
\cite{karrai_photonics:cooling_2006,arcizet_high-sensitivity_2006,regal_measuring_2008,groblacher_demonstration_2009,park_resolved-sideband_2009,schliesser_resolved-sideband_2009,aspelmeyer_quantum_2010},
through  
electrostatic effects
\cite{poncharal_electrostatic_1999,ilic_attogram_2004,lahaye_approachingquantum_2004,hertzberg_back-action-evading_2010},
mechanical \cite{garcia-sanchez_mechanical_2007}, or even
field emission in 
the case of a nano-tube \cite{purcell_tuning_2002}) were used. Most of these
do use linear transduction, but 
whether this is the optimal measurement procedure is an open question. 

The truly fundamental lowest (but achievable) bound on the mass
sensitivity is a 
function of the quantum state 
of the resonator, and optimized over all possible measurement procedures.  It
will be calculated below using quantum parameter estimation theory, which
leads to the ultimate limit of sensitivity, the quantum Cram\'er-Rao bound
\cite{Braunstein94}. It becomes relevant once all other
limitations such as technical noise, adsorption-desorption noise,
momentum exchange noise, etc.~have been
eliminated \cite{ekinci_ultimate_2004}.  I will even assume a harmonic
oscillator without any dissipation (and thus decoherence effects), as mixed
states can only decrease the ultimate sensitivity compared to the pure
states from which they are mixed \cite{braun_parameter_2010}.  Nevertheless,
the  bounds I calculate are attainable {\em in principle} if the idealized
conditions are met, and therefore set an important benchmark to which 
the performance of existing 
sensors should be compared to.  As a guide to further improving the sensitivity
of mass-sensing 
using quantum-engineered states of a nano-oscillator, I determine the
optimal quantum state for a given maximum number of 
excitation quanta in the oscillator.\\

\section{Quantum parameter estimation theory}
For small enough excitation amplitudes, the nano-mechanical resonator can
be modelled as a harmonic oscillator with mass $M$ and effective spring
constant $D$ \cite{jensen_atomic-resolution_2008}, 
resonance frequency $\omega=\sqrt{D/M}$, and hamiltonian  
$H_\omega=\hbar\omega (a^\dagger_\omega a_\omega+\frac{1}{2})$ with the
usual raising (lowering) operators $a^\dagger_\omega$ 
($a_\omega$). If a small mass $\delta M$ is added to the oscillator, its
frequency changes to $\tilde{\omega}=\omega(1-\epsilon)$ with
$\epsilon=(1/2)\delta M/M$, and we obtain the new hamiltonian
$H_{\tilde{\omega}}$  
from  $H_\omega$ by replacing $\omega\to\tilde{\omega}$ everywhere. An
arbitrary initial quantum state $\rho_0$ is thus propagated to
$\rho(\omega,t)=U(\omega,t)\rho_0 U^\dagger(\omega,t)$ (or
$\rho(\tilde{\omega},t)$, respectively),  if no mass (or the mass $\delta M$)
is adsorbed at $t=0$, where $U(\omega,t)=\exp(-\ri H_\omega
t/\hbar)$. Note that this assumes that the energy of the oscillator is
conserved in the adsorption process, i.e.~the additional mass is
deposited with zero differential speed onto the oscillator. The
distinguishability of the two states $\rho(\omega,t)$ and
$\rho(\tilde{\omega},t)$ determines the smallest $\delta M$ that can be
measured. In general, for any density matrix $\rho(x)$ that depends on some
parameter $x$, the smallest $\delta x$
that can be resolved from $N$ measurements of an observable $A$ (starting
always from an identically prepared state) is given by
\cite{Braunstein94}   
\begin{equation} \label{dxA}
\delta x=\frac{\langle \delta
  A^2\rangle_x^{1/2}}{\sqrt{N}|\frac{\partial}{\partial x}\langle A\rangle_x|}\,. 
\end{equation}
It has the interpretation of the uncertainty of $A$ in
state $\rho(x)$ as judged by $N$ measurements, renormalized by the ``speed''
by which the mean value of $A$ changes as function of $x$. In other words,
$x$ has to change by an amount that moves the average value of $A$ by at
least its uncertainty. Optimizing over all possible measurements
leads to the quantum Cram\'er-Rao bound
\cite{Braunstein94},    
\begin{equation} \label{dx}
\delta x\ge\delta x_{\rm min}\equiv
\frac{1}{2\sqrt{N}\frac{d_{\rm
      Bures}(\rho(x),\rho(x+dx))}{dx}}\,,  
\end{equation}
 where $d_{\rm Bures}(\rho(x),\rho(x+dx))$ is the Bures distance
 between $\rho(x)$ and $\rho(x+dx)$ (also called Fisher information),
 defined as $d_{\rm 
 Bures}(\rho_1,\rho_2)=\sqrt{2}\sqrt{1-\sqrt{F(\rho_1,\rho_2)}}$ through the
 fidelity
 $\sqrt{F(\rho_1,\rho_2)}=\tr((\rho_1^{1/2}\rho_2\rho_1^{1/2})^{1/2})$.  
Thus, in our case, we obtain the minimal measurable mass $\delta M_{\rm
 min}$ by evaluating the
 Bures distance between $\rho(\omega,t)$ and $\rho(\tilde{\omega},t)$ in the
 limit $\epsilon\to 0$.  It is important to note that (\ref{dx}) is, in the
 limit of large $N$, an {\em achievable} lower bound \cite{Braunstein94}. It
 does take into account effects such as back-action and the quantum noise of
 the system.\\

\section{Pure states}
In the case of two pure states, we have simply
$F(|\psi\rangle\langle\psi|,|\phi\rangle\langle\phi|)^{1/2}=|\langle\psi|\phi\rangle|$.
Starting from an initial state $|\psi(0)\rangle=\sum_{n=0}^\infty
c_n|n\rangle_\omega$, we have the overlap at time $t$,     
\begin{eqnarray} \label{pp1}
\langle\psi(t)|\tilde{\psi}(t)\rangle&=&
\sum_{n,m,k}c_n^*c_k\re^{-i(\tilde{E}_m-E_n)t/\hbar}\nonumber\\
&&\times R_{\tilde{\omega}\omega}(m,k)R_{\tilde{\omega}\omega}(m,n)\,,    
\end{eqnarray}
where $R_{\tilde{\omega}\omega}(m,n)={_{\tilde{\omega}}}\langle
m|n\rangle_\omega=R_{\omega\tilde{\omega}}(n,m)$ denotes the overlap matrix
element between energy eigenstates of the two oscillators with frequency
$\tilde{\omega}$ and $\omega$, and the coefficients $c_n^*$, $c_k$ are
expressed in the energy eigenbasis of the unperturbed 
oscillator. They are \cite{smith_overlap_1969}
\begin{eqnarray} \label{R}
R_{\tilde{\omega}\omega}(m,n)&=&(2^{-(m+n)}qm!n!)^{1/2}\\
&&\times\sum_{r=0,1}^{[m,n]}\frac{(2q)^r}{r!}\frac{y^{(m+n-2r)/2}(-1)^{(m-r)/2}}{(\frac{1}{2}(n-r))!(\frac{1}{2}(m-r))!}\,, \nonumber 
\end{eqnarray}
if $m,n$ are both even or both odd (otherwise
$R_{\tilde{\omega}\omega}(m,n)=0$), and $[m,n]$ 
denotes the smaller of the two integers $m,n$. The sum over $r$ runs over
even (odd) integers for $[m,n]$ even (odd), respectively, and
$y=(\omega-\tilde{\omega})/(\omega+\tilde{\omega})$,
$q=2(\omega\tilde{\omega})^{1/2}/(\omega+\tilde{\omega})$. We need
$|\langle\psi(t)|\tilde{\psi}(t)\rangle|$ to second order in $\epsilon$. We
find $F=1+\epsilon^2f(\{c_m\},t)$ with 
\begin{eqnarray}
f(\{c_m\},t)&=&\Big[
\sum_{m=0}^\infty\big\{\frac{1}{2}\sqrt{(m+1)(m+2)}\label{f}\\
&\times&\Im(c_mc_{m+2}^*(\re^{2\ri\tau}-1))+\tau m|c_m|^2\big\}
\Big]^2\nonumber\\
&-&\sum_{m=0}^\infty\Big\{\left(\frac{(m^2+m+1)\sin^2\tau}{2}+m^2\tau^2\right) |c_m|^2\nonumber\\
&-&\sqrt{(m+1)^3(m+2)}\tau\, \Im((1-\re^{2\ri   \tau})c_{m+2}^*c_m)\nonumber\\
&-&\frac{1}{8}\sqrt{(m+1)(m+2)(m+3)(m+4)}\nonumber\\
&\times&\Re((1-\re^{2\ri\tau})^2c_mc_{m+4}^*)
\Big\}\,\nonumber
\end{eqnarray}
and $\tau=\omega t$.
Inserting $F$ in $d_{\rm Bures}$, we find immediately $d_{\rm 
  Bures}(|\psi(t)\rangle\langle\psi(t)|,|\tilde{\psi}(t)\rangle\langle\tilde{\psi}(t)|)=\epsilon
  |f(\{c_m\},t)|^{1/2}$, and thus 
\begin{equation} \label{dmm}
\frac{\delta M_{\rm min}}{M}= \frac{1}{\sqrt{N}|f(\{c_m\},t)|^{1/2}}\,.
\end{equation}
Eq.(\ref{dmm}) together with (\ref{f}) constitutes the central result of
this report which we now explore for particular cases. 

\subsection{Fock state}
For $|\psi(0)\rangle=|n\rangle$, we have $f=-(1/2)(n^2+n+1)\sin^2\tau \equiv f_n^{\rm Fock}$
for all $n\ge 0$. The largest absolute value is achieved for
$\tau=\pi/2(\mod 2\pi)$, and leads to  
\begin{equation} \label{fFock}
\frac{\delta M_{\rm min}}{M}=
\sqrt{\frac{2}{N}}\frac{1}{\sqrt{n^2+n+1}}\sim\sqrt{\frac{2}{N}}\frac{1}{n}\mbox{
  for }n\gg 1\,.
\end{equation}
Thus, one can measure, at least in principle, arbitrarily small masses
within the same fixed time interval by
increasing the excitation of the harmonic oscillator.  In reality, of
course, non-harmonicities will start to arise at some level of excitation
and the present analysis will then have to be extended to a more complicated
hamiltonian \cite{Dunn10}.  The ground state  $n=0$ of the harmonic
oscillator allows to measure a mass which, for a single readout, can
be of the order of the mass of the oscillator itself, $\delta
M/M\ge\sqrt{2/N}/|\sin\tau|$.  Increasing $\tau$ does not help beyond
$\tau=\pi/2$, as $f_n^{\rm Fock}$ is periodic in $\tau$.

\subsection{Little Schr\"odinger cat states}
Given that (\ref{f}) depends on coherences between states $|n\rangle$,
$|n+2\rangle $, and $|n+4\rangle$, one might wonder whether the precision
could be 
increased further by using superpositions of these states. The state
$|\psi(0)\rangle=(|n\rangle+|n+2\rangle)/\sqrt{2}\equiv|\psi_{S1}\rangle$
leads to  
\begin{equation} \label{f2}
f=\frac{1}{16}\big((n+1)(n+2)\sin^2(2\tau)-8(n^2+3n+4)\sin^2\tau\big)-\tau^2\,.  
\end{equation}
 The maximum of the periodic term is again achieved for $\tau=\pi/2\,(\mod
2\pi)$. For fixed $\tau\ne k\pi$ ($k\in\mathbb{N}$) and $n\gg \tau$, this
term dominates and leads to 
$\delta M_{\rm min}/M\simeq\sqrt{2/N}/n$, just as for the Fock
state. However, for fixed 
$n$, we can get an arbitrarily small $\delta M/M$ by increasing $\tau$, as
the last term in (\ref{f2}) leads to $\delta M_{\rm
  min}/M\simeq\sqrt{2/N}/\tau$ for  
$\tau\gg n$. Note that this improvement is beyond the usual factor $1/\sqrt{t}$
from increasing the measurement time.  Indeed, the sensitivity
$\delta M \sqrt{t}$ in (g/$\sqrt{\mbox{Hz}}$) still improves as $\propto
1/\sqrt{t}$. 

The state
$|\psi(0)\rangle=(|n\rangle+|n+4\rangle)/\sqrt{2}\equiv |\psi_{S2}\rangle$
gives  
\begin{eqnarray} \label{f4}
f&=&-\frac{1}{4}\Big(2(n^2+5n+11)\sin^2\tau\\
&&+\sqrt{(n+1)(n+2)(n+3)(n+4)}\cos(2\tau)\sin^2\tau\Big)\nonumber\\
&&-4\tau^2\,.  \nonumber
\end{eqnarray}
The maximum of the periodic terms (relevant for fixed $\tau$ and
$n\gg\tau\gtrsim 1$) is close to $\tau=\pi/3$ with $|f|\simeq 9n^2/32$, which
gives a $\delta M_{\rm min}$  
33\% larger than for a Fock state with $n$ excitations. For fixed
$n$ and $\tau\gg n$, the factor 4 in front of $\tau^2$ in (\ref{f4})
reduces $\delta M_{\rm min}/M$ by a factor 2 compared to 
$|\psi_{S1}\rangle$.  

\subsection{Coherent state}
For a coherent state
$|\alpha\rangle=\exp(-|\alpha|^2/2)\sum_{n=0}^\infty(\alpha^n/\sqrt{n!})|n\rangle$, 
and $\alpha\in\mathbb{R}$, we have 
\begin{equation} \label{fcoh}
f=-\left(\frac{1}{2}+\alpha^2\right)\sin^2\tau-\alpha^2\tau\left(\tau+\sin
(2\tau)\right)\,.\nonumber\,\,\,\,\,\, (10)
\end{equation}
For fixed $\tau\gg 1$ and $\alpha^2\gg 1$, we find $f\simeq
-\alpha^2\tau^2$, and hence $\delta
M_{\rm min}/M\simeq 1/(\sqrt{N}\alpha\tau)=1/(\sqrt{N\langle n\rangle}\tau)$.  
Thus, for a 
coherent state, the sensitivity scales as the inverse square root of the 
(average) number of 
excitations in the oscillator, and one can, at least in principle, resolve
arbitrarily small masses. We see again that $\delta M_{\rm
  min}$ reduces faster than $1/\sqrt{t}$ with measurement time. Compared to
a Fock state with $n=\alpha^2$, there 
is  
a factor $\sqrt{\langle n\rangle/2 }$ penalty in the scaling with $\langle
n\rangle$ (but one gains a factor $1/\tau$). For $\alpha=0$ we find
$f=-\sin^2\tau/2$ 
which leads back to the result for the Fock state $n=0$. ``\\

\subsection{Optimal state}
\begin{figure}
\epsfig{file=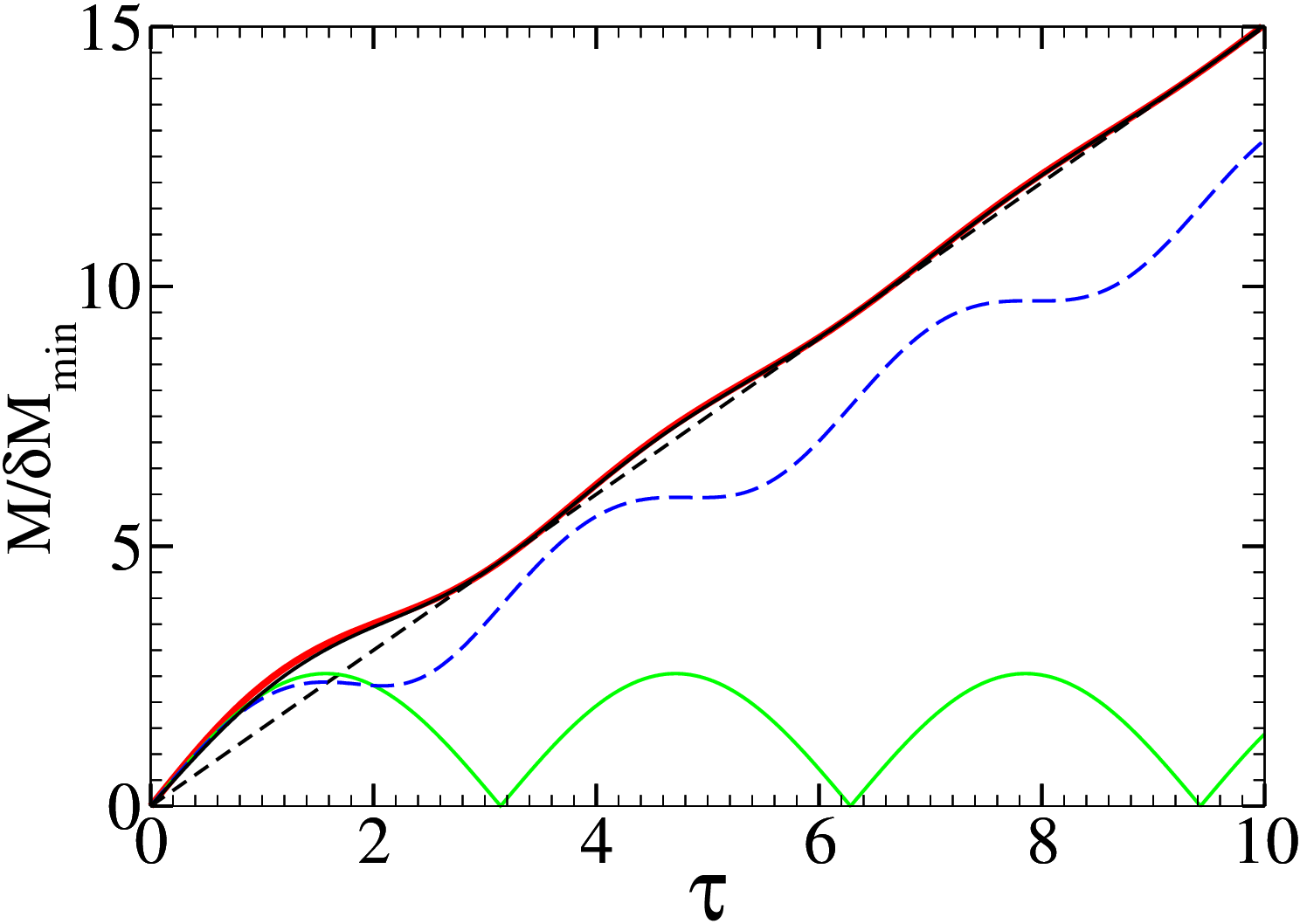,width=6cm,angle=0}
\caption{Inverse minimal measurable mass $M/\delta M_{\rm min}$ (for $N=1$)
  as function of $\tau$ for selected pure states. Green line: Fock state
  $|n=3\rangle$; Full black 
  line: ON state $|\psi_{\rm ON}\rangle=(|0\rangle+|3\rangle)/\sqrt{2}$;
  Dashed black line: 
  asymptotic behavior $3\tau/2 $ for $|\psi_{\rm ON}\rangle$; Red line:
  optimal state with at most   $L=3$ 
  quanta in the oscillator; Dashed blue line: Coherent state with the same {\em
  average} number of excitations as the ON state ($\langle
  n\rangle=3/2$). }\label{fig1}         
\end{figure}

What is the
best sensitivity that can be achieved for a given maximum number $L$ of
excitations and fixed measurement time?
From (\ref{f}) we see that for $\tau\gg 1$, the terms quadratic in $t$
dominate and give simply $|f|=\tau^2(\langle n^2\rangle-\langle
n\rangle^2)$. Hence, in this case the optimal pure state is the one which
maximizes the excitation number fluctuations. One easily shows that this
state has the 
form of an ``ON'' state (half a ``NOON'' state \cite{Sanders89}),
$|\psi_{\rm 
 ON}\rangle=(|0\rangle+\re^{\ri\varphi}|L\rangle)/\sqrt{2}$, where 
$\varphi$ is an irrelevant phase which we will choose equal zero.  
It leads to $|f_{\rm ON}|\equiv|f(\psi_{\rm ON},\tau)|\simeq\tau^2L^2/4$,
and 
thus a minimal mass $\delta M_{\rm 
  min}/M=2/(\sqrt{N}\tau L)$.  Fig.\ref{fig1} shows a comparison of the
(inverse) minimal mass for $\psi_{\rm ON}$ with the true minimal mass for
given $\tau$ and the same $L$, obtained by numerically maximizing
$|f|$, for  
$L=3$.  We see that  $|f_{\rm ON}|$ approximates the best possible $|f|$
very well, even for $\tau\sim 1$.  For $\tau=k\pi$, $k\in\mathbb{N}$,
$|f_{\rm ON}|$ gives in fact the exact result, as is obvious from 
(\ref{f}). Fig.\ref{fig1} also shows the result for a coherent state with
the same {\em average} number of excitations as the ON state, $\langle
n\rangle=3/2$. It leads for small $\tau$ to comparable sensitivity as the
optimal state with 
$L=3$.   At $\tau=\pi/2$, the optimal pure state with
$L=3$ 
allows still a reduction of $\delta M_{\rm 
  min}/M$ by $\sim 4$\% compared to $\psi_{\rm ON}$, and by $\sim 18$\%
compared to the Fock 
state with the same $L$.

\begin{figure}
\epsfig{file=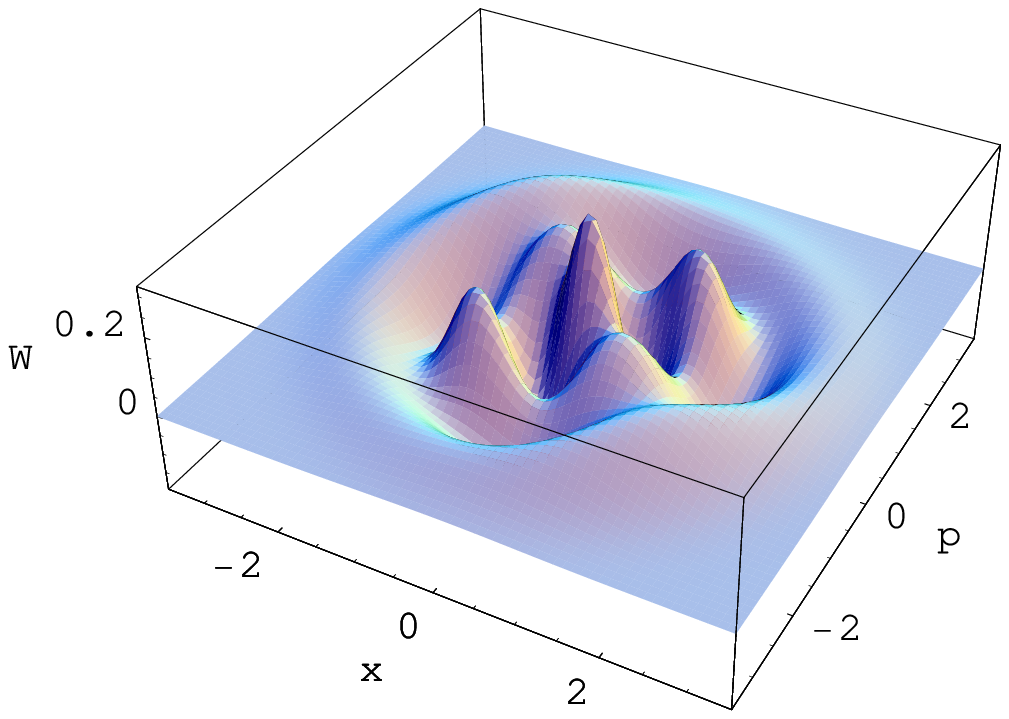,width=6cm,angle=0}
\caption{Wigner function of the optimal
  initial state for $L=4$, $\tau=\pi/2$}\label{fig2}         
\end{figure}

Fig.\ref{fig2} shows the
  Wigner
function, defined for a pure state $|\psi\rangle$ by \cite{GardinerZoller04}
\begin{equation} \label{W}
W(x,p)=\frac{1}{\pi}\int_{-\infty}^\infty
dy\psi^*(x-y)\psi(x+y)\re^{-2\ri y p}\,,  
\end{equation}
(with all lengths in units of the oscillator length
$x_0=\hbar^{1/2}/(DM)^{1/4}$ and 
$p$ in units of $\hbar/x_0$), for the optimal state for $L=4$ and
$\tau=\pi/2$, $|\psi_{\rm opt}\rangle\simeq
(-0.62057+0.0305\ri)|0\rangle
-(0.00059+0.01198\ri)|2\rangle+(0.78252-0.038852\ri)|4\rangle$. $|\psi_{\rm 
 ON}\rangle$ and  $|\psi_{\rm opt}\rangle$ have very similar Wigner
functions, characterized by four lobes in azimuthal direction which
guarantee minimal phase uncertainty, as is to be expected from the
requirement of minimal noise and maximum uncertainty in the number of
excitations.  Rotations of $|\psi_{\rm opt}\rangle$ through evolution
with the unperturbed hamiltonian before the adsorption of mass clearly leave
$\delta M_{\rm min}$ invariant.
\begin{figure}
\epsfig{file=invMinMassOfTandX.eps,width=6cm,angle=0}
\caption{Inverse minimal measurable mass $M/\delta M_{\rm min}$ in a thermal
  state as function of dimensionless time $\tau=\omega t$ for inverse
  dimensionless temperatures     $z=\hbar\omega/(k_BT)=0.2,0.5,1.0,2.0,5.0$
  and 10.0 (red, orange, yellow, green, blue, and 
  dashed purple line, respectively). Inset: $M/\delta M_{\rm min}$ as
  function of $z$ for $\tau=\pi/2$.}\label{fig3}         
\end{figure}
It is an open question how to experimentally realize the optimal
state. In \cite{oconnell_quantum_2010}, non-trivial quantum states
were obtained by coupling the harmonic oscillator to a
super-conducting qubit whose quantum state can be readily
manipulated. Other schemes for engineering the quantum state of a
nano-mechanical harmonic oscillator were proposed in \cite{Bose97}.\\

Among the pure states considered, the coherent states certainly come closest
to the typical experimental situation, where the oscillator is cooled to low
temperature and driven on resonance. Inserting typical numbers for
micromachined resonators,
$M=10^{-16}$g \cite{giscard_quantum_2009}, $\omega=$1GHz, an evolution time
$\tau=10^6$, and an excitation with $\langle n\rangle 
\sim 10^{10}$ quanta (driving energy $E_d=10^{-15}$J in
\cite{giscard_quantum_2009}), we find $\delta M_{\rm min}\simeq 10^{-27}{\rm
  g}/\sqrt{N}$, or roughly the mass of an electron for a single readout,
$N=1$. Higher masses (e.g. $M\sim 10^{-14}g$ in
\cite{yang_zeptogram-scale_2006}) give proportionally higher $\delta M_{\rm
  min}$, everything else equal.
In assuming $\tau=10^6$, we have made a pessimistic estimate in the
sense of using the shortest sensing 
time allowed by the inverse bandwidth 1 kHz. Measuring during longer times
decreases $\delta M_{\rm min}$ 
further.  On the other hand, $\tau$ might be limited  by
decoherence and
dissipation, such that for later times another (mixed) quantum state becomes
relevant. These effects are, at least in principle, avoidable, and
in such a 
highly idealized situation (which is, however, relevant as utlimate
achievable goal), $\tau$ in eq.(\ref{f}) is given by the measurement
time. 
The mass $\delta M_{\rm min}\simeq 10^{-27}{\rm
  g}$ agrees  with the prediction of \cite{giscard_quantum_2009}, 
but the agreement appears to be a coincidence: The result in
\cite{giscard_quantum_2009}, based purely on noise considerations, still
decreases as $1/\sqrt{Q}$ with the quality 
$Q$ of the  
resonator, whereas (\ref{fcoh}) is independent of $Q$, taken as infinity in
the present analysis. Also, while in the regime $\tau\gg \alpha$ 
relevant for the above numbers ($\alpha=10^5$) both $\delta M_{\rm 
  min}/M$ and the 
result in \cite{giscard_quantum_2009} scale as $1/\sqrt{\langle n\rangle}$,
\cite{giscard_quantum_2009} predicts a proportionality to $1/\sqrt{\tau}$ if
one identifies the inverse bandwidth $1/\Delta f$ with $t$, 
instead of the $1/\tau$ behavior that follows from (\ref{fcoh}). 

Carbon nanotube resonators have typically much smaller masses than
micro-engineered ones (of order $M\simeq 10^{-18}$g
\cite{jensen_atomic-resolution_2008}) with comparable resonance
frequency ($\omega=2\pi\times 328.5$MHz in
\cite{jensen_atomic-resolution_2008}), and can therefore resolve in principle
even smaller masses. Assuming a coherent state with oscillation amplitude of
about 10nm for the carbon 
nanotube resonator in \cite{jensen_atomic-resolution_2008} and a sampling
time of 100ms, $\delta M_{\rm min}$ according to (\ref{fcoh}) is of
the order of a thousandth of an electron mass.

\section{Mixed states}
In general, due to the joint-convexity of the Bures distance, mixed
states do not allow better sensitivities than the 
pure states from which they are mixed, as long as the weights are
independent of the parameter to be measured and the evolution is
linear in the state \cite{braun_parameter_2010}. Evaluating the
Bures distance 
between two mixed states is much more difficult 
than for pure states. Nevertheless, one can evaluate the Bures distance
numerically.  One may also obtain an upper bound on
$d_{\rm Bures}$ using the joint-convexity of $d_{\rm Bures}$, which leads to
a (typically non-achievable) lower 
bound on $\delta M_{\rm min}$.  An achievable upper bound on $\delta
M_{\rm min}$ can be found by considering a particular measurement 
$A$ in  (\ref{dxA}). 

As an example, consider the thermal state $\rho=\sum_{n=0}^\infty
p_n|n\rangle\langle n|$, with $p_n=\re^{-n
  z}(1-\re^{-z})$, $z=\beta\hbar\omega$,  
$\beta=1/k_B T$ the inverse temperature, and $k_B$ the Boltzmann constant. 
Using the invariance of the thermal state under the time evolution governed by
$H_\omega$ and the joint convexity of $d_{\rm Bures}(\rho_1, \rho_2)$, one
shows easily that $d_{\rm Bures}/dx\le \sum_n p_n|f_n^{\rm Fock}|^{1/2}\le
|\sin\tau|/(\sqrt{2}(1-\exp(-z))$.  For $z\to \infty$, this bound coincides 
with the exact result for the groundstate $n=0$. We may choose a measurement
of the width $A=x^2$ as a way of measuring the change of mass. 
Thermal average $\langle
x^2\rangle=(\hbar/(2\sqrt{DM}))\coth(z/2)$ and fluctuations $\langle
\delta(x^2)\rangle =(\langle x^4\rangle-\langle
x^2\rangle^2)^{1/2}=(\hbar/\sqrt{2 DM})\coth(z/2)$ give an
achievable upper bound $\delta M_{\rm min}/M\le 2\sqrt{2/N}\sinh z/(\sinh
z-z)$.  For $z\to\infty$ this bound is only a factor 2 above the best
possible value for
the groundstate $n=0$, whereas for $z\to 0$ 
the bound diverges.  Fig.\ref{fig3} shows the exact $M/\delta M_{\rm min}$
obtained by evaluating the Bures distance numerically. We see that $\delta 
M_{\rm min}/M$ is periodic in $\tau$, just as for the ground
state. Increasing the temperature helps, as higher Fock states start to
contribute, but at most a factor $\sqrt{2}$ can be gained, and $\delta
M_{\rm min}$ remains bounded by the mass of the oscillator itself for all
temperatures. 

\section{Conclusions}
In summary, I have calculated the smallest measurable adsorbed mass
$\delta M_{\rm min}$ on 
a nano-mechanical harmonic resonator in an arbitrary pure state, based on
the fundamental quantum Cram\'er-Rao bound. The 
analysis shows that a coherent state allows to achieve a $\delta M_{\rm
  min}/M$ that scales for fixed measurement time as the 
inverse square root of 
the average number of excitations.  For a given maximum number $n$ of
excitations, I found the optimal quantum state, which for large $\tau$ is
$|\psi_{\rm ON}\rangle=(|0\rangle+|n\rangle)/\sqrt{2}$, with 
a 
sensitivity that scales as $1/n$. For $\tau\ne k\pi$, $k\in\mathbb{N}$, the
sensitivity can be further enhanced. Even 
with a coherent state of a carbon nanotube resonator
\cite{jensen_atomic-resolution_2008}, the 
smallest resolvable mass should be of the order of a thousandth of an
electron mass. If two more orders of magnitude could be gained (say by
increasing $\tau$ and the number of excitations), the regime
could be reached where one can weigh the relativistic mass change due to
the formation of a chemical bond or the absorption of a photon (energies of
order 1 eV).

%\acknowledgments

\bibliography{../mybibs_bt}

\end{document}